\documentclass[showpacs,aps,pra,amsmath,amssymb,twocolumn]{revtex4}
\usepackage{graphicx}
\usepackage{dcolumn}
\usepackage{bm,color}
\usepackage{txfonts}
\usepackage{amsmath,epsfig}
\usepackage{epstopdf}
\newcommand{\eq}{Eq.~}
\newcommand{\eqs}{Eqs.~}
\newcommand{\fig}{Fig.~}
\newcommand{\figs}{Figs.~}
\newcommand{\cf} {cf.~}
\newcommand{\ug} {\!=\!}

\newcommand{\piu} {\!+\!}
\newcommand{\meno} {\!-\!}

\newcommand{\eg} {e.g.~}

\newcommand{\rrefs} {Refs.~}

\begin{document}

\author{F. Lombardo$^1$,  F. Ciccarello$^2$ and G. M. Palma$^2$}
\affiliation{$^1$Dipartimento di Fisica e Chimica, Universit$\grave{a}$  degli Studi di Palermo, via Archirafi 36, I-90123 Palermo, Italy\\
$^2$NEST, Istituto Nanoscienze-CNR and Dipartimento di Fisica e Chimica, Universit$\grave{a}$  degli Studi di Palermo, via Archirafi 36, I-90123 Palermo, Italy}

% 03.65.Yz Decoherence; open systems; quantum statistical methods
% 37.30.+i: atoms, moleculs etc in cavities

\title{Photon localization versus population trapping in a coupled-cavity array}

\date{\today}
\begin{abstract}
We consider a coupled-cavity array (CCA), where one cavity interacts with a two-level atom under the rotating-wave approximation. We investigate the excitation transport dynamics across the array, which arises in the atom's emission process into the CCA vacuum. Due to the known formation of atom-photon bound states, partial field localization and atomic population trapping in general take place. We study the functional dependance on the coupling strength of these two phenomena and show that the threshold values beyond which they become significant are different. As the coupling strength grows from zero, field localization is exhibited first.
 \end{abstract}
\maketitle

\noindent

\section{Introduction}

Coupled-cavity arrays (CCA)  typically consist of an arrangement of low-loss cavities with nearest-neighbor coupling allowing photon hopping between neighbouring cavities. In turn, each cavity may interact with one or more atoms (or atom-like systems). Progress in the fabrication techniques make such systems experimentally accessible or nearly so in the immediate future \cite{plenio}. At the same time their extremely rich physics is triggering a strong attention to the behaviour of these objects from a fundamental as well as applicative viewpoint due to their potential to work as an effective platform to carry out quantum information processing and photonics tasks. {In particular, an interesting and rich dynamics characterises the propagation of initially localized excitations along the array \cite{excitation}. In this respect, the simplest -- yet very interesting -- scenario is the propagation of excitation in single-atom arrays, where only  one cavity of the CCA is coupled to a single two-level atom as sketched in \fig1(a).} 
Most, if not all, of the attention to such setup has been focused on single- or multi-photon scattering \cite{scattering}, where one or more photons impinge on the initially unexcited atom.

The single-atom CCA -- under rotating-wave approximation and in the thermodynamic limit of an infinite number of cavities, is a known instance of the general Fano-Anderson (or Friedrichs-Lee) model, where a two-level system (TLS) is coupled to a {\it finite} band of bath modes with a constant coupling strength \cite{mahan, pronko}. When the TLS frequency lies at the middle of the band, the resulting dressed system features a continuous band (coinciding with the bare bath band) and two symmetrical out-of-band discrete levels. While the former has associated unbound stationary states (corresponding to photon scattering states in our case), either discrete level corresponds to a TLS-bath bound state. For a single-atom CCA, each bound state is localized around the atom's position on the lattice (i.e., the cavity to which it is coupled), the localization length being a decreasing function of the coupling strength. In the limit where this is is very high compared to the band width, the pair of BSs reduce to the pair of dressed states of the well-known Jaynes-Cummings model in the single-excitation sector. Some aspects of such bound states have been recently discussed mostly in the framework of photon scattering problems \cite{sun,cinesi, longo,tureci}. While in these works one or more photons impinge on the initially unexcited atom, here we will focus on what can be regarded as a sort of inverse process, namely the atom's emission when the CCA is initially in the vacuum state. Still, a major goal of ours is to characterize the essential features of the resulting photon transport.

Owing to the presence of the atom-photon bound states, the atom is in general unable to eventually release the entire amount of initial excitation to the field and thus exhibits fractional decay \cite{gaveau,petrosky}. Such {\it population trapping} manifests in the form of a residual oscillatory behavior of the excited state population at long times. If population trapping takes place then, clearly, (partial) {\it photon localization} must occur. In the high-coupling-strength limit, the field remains confined within the cavity coupled to the atom, which gives rise to a full, time-continuous, atom-field energy swapping (the aforementioned stationary oscillations reducing to the standard vacuum Rabi oscillations). Based on this, one naturally wonders how such energy exchange takes place at lower coupling strengths between the atom and the localized fraction of the field. Since photon localization and population trapping are both due to the emergence of bound states, one might expect such phenomena to arise simultaneously as the coupling rate grows from zero. Instead, we show here that this is not the case. Specifically, we highlight the existence of a regime of intermediate coupling strengths such that significant field localization can occur with no appreciable population trapping.
In such cases, thereby, field localisation in fact is not accompanied by appreciable fractional atomic decay.  At long enough times, in this regime the localized fraction of the photon wave function undergoes time modulation in intensity. Unlike in the strong-coupling limit, such modulation is not connected with atomic excitation/emission processes (the atom having fully decayed to the ground state). Rather, it reflects a mere time-continuous redistribution of light among the cavities next to the atom.

This paper is organised as follows. In Section \ref{sec-h}, we introduce the Hamiltonian model. In Section \ref{sec-ss}, we derive the system's stationary states, both unbound and bound ones through the unifying Green function approach. We use these in Section \ref{emission-sec} with the aim to study the atomic emission process, focusing on the time behaviour of both the atom population and photon probability distribution function. This way, we identify three main regimes. In Section \ref{study}, we analyse in detail the regime in which one observes field localization with no population trapping. Such dynamics is linked to the properties of the atom-photon bound states. Finally, in Section \ref{concl} we draw our conclusions. Some technical details are given in the Appendixes.

\section{Hamiltonian model}\label{sec-h}

The system under study consists of an array of $N\!\gg\!1$ single-mode identical cavities and a two-level atom -- whose ground and excited state are denoted by $|g\rangle$ and $|e\rangle$, respectively -- which is resonantly coupled to one of the cavities. By engineering the cavity array in such a way that the field modes exhibit spatial overlap 
photon hopping can occur between nearest-neighbor cavities \cite{plenio}. A sketch of the entire setting is shown in \fig1(a).
The Hamiltonian reads (we set $\hbar\ug1$ throughout)
\begin{equation}
\hat H=\hat H_0+\hat H_1\,\label{H}
\end{equation}
with
\begin{eqnarray}
\hat H_0&=&J \sum_{x=-N/2}^{N/2-1}  (\hat{a}_{x+1}^\dag\hat{a}_x+{\rm h.c.})\,,\label{H0}\\
\hat H_1&=& g (\hat {\sigma}_+\hat{a}_{0}+{\rm h.c.})\,,\label{H1}
\end{eqnarray}
where $\hat \sigma_+\ug\hat \sigma_-^\dag \ug|e\rangle\langle g|$, $\hat a_x$ ($\hat a_x^\dag$) annihilates (creates) a photon at the $x$th cavity and index $0$ labels the cavity which the atom is coupled to. The Hamiltonian is fully specified by the two characteristic rates $J$ and $g$, namely the hopping and atom-photon coupling rates, respectively. In deriving Hamiltonian (\ref{H}), we have assumed that the atom is on {\it resonance} with the $0$th cavity, i.e., the atom and cavity frequencies coincide (let us call $\omega_0$ their common value). This makes the dynamics independent of $\omega_0$, which allows us to set $\omega_0\ug0$. Moreover, we assume {\it cyclic boundary conditions} for the field, i.e., ${\bf\hat{a}_{N/2}\!\equiv\!\hat{a}_{-N/2}}$, since we will be interested in the emission problem of an atom into an infinitely long array.

The form of (\ref{H0}) and (\ref{H1}) entails conservation of the total number of excitations, i.e.,  $[\hat H,\hat\sigma_{+}\hat\sigma_-\piu \sum_x\hat{a}_x^\dag\hat{a}_x]\ug0$. In the following we will restrict our attention to atomic emission in vacuum, with the atom initially in its excited state and no photons present. With such initial conditions the dynamics takes place within the one-excitation subspace spanned  by $\{|e\rangle|{\rm vac}\rangle, |g\rangle\hat{a}_{-N/2}^\dag|{\rm vac}\rangle,...,|g\rangle\hat{a}_{N/2-1}^\dag|{\rm vac}\rangle\}$, where $|{\rm vac}\rangle$ is the field vacuum state. It is immediate to see that the present system is effectively equivalent to the network sketched in \fig1(b) consisting of a linear chain with a stub connected to the central site. This shows that the dynamics can be mapped into that of an atom-free coupled-cavity network, where the atom is in fact replaced by a further effective cavity.
For the sake of simplicity, from now on we use the coincise notation $|e\rangle\!\equiv\!|e\rangle|{\rm vac}\rangle$ and $|x\rangle\!\equiv\!|g\rangle\hat{a}_x^\dag|{\rm vac}\rangle$.
\begin{figure}
\includegraphics[width=0.48\textwidth]{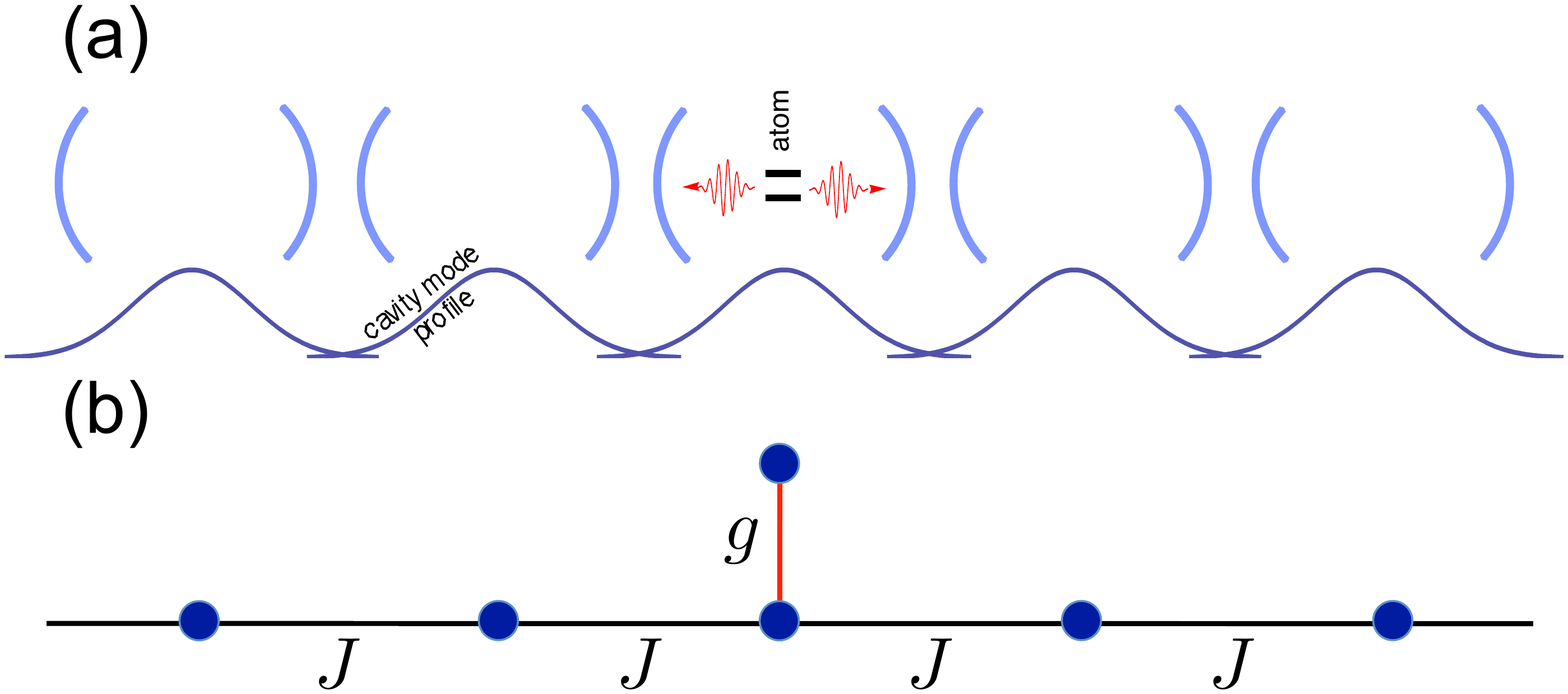}
\caption{(Color online){(a) Sketch of the setup: a large array of single-mode cavities, where one cavity is coupled to an initially excited two-level atom. Photon hopping can occur between two nearest-neighbour cavities owing to the spatial overlap of the corresponding modes' profile. (b) Equivalent network.}  
 \label{Fig1}}
\end{figure} 

$\hat H_0$ can be expressed in the well-known diagonal form
\begin{eqnarray}
\hat H_0&=&\sum_{k} \omega_k \,|\varphi_k\rangle\langle \varphi_k|\,
\end{eqnarray}
with
\begin{eqnarray}
k&=&\frac{2\pi n}{N}\label{k}\,\,\,\,\,\,\,\,(n=-N/2,-N/2+1,...,N/2-1)\,\,,\\
\omega_k&=&2J\cos k\label{omegak}\,\\
 |\varphi_k\rangle&=&\frac{1}{\sqrt{N}}\sum_{x=-N/2}^{N/2-1} e^{i k x}|x\rangle\,.\label{phik}
\end{eqnarray}
\eqs(\ref{omegak}) and (\ref{phik}) represent, respectively, the free photon dispersion law and the associated field normal modes. The possible (free) photon energies fall in the frequency range $\omega_k\!\in\![-2J,2J]$, which becomes a continuous band in the thermodynamic limit $N\!\rightarrow\!\infty$.

In the basis $\{|e\rangle,\{|\varphi_k\rangle\}\}$ the Hamiltonian (\ref{H}) takes the form
\begin{eqnarray}
\hat H&=&\sum_{k} \omega_k \,|\varphi_k\rangle\langle \varphi_k|+\sum_k\frac{g}{\sqrt{N}}\left( |e\rangle\langle\varphi_k|+{\rm H.c.}\right)\,,
\end{eqnarray}
which shows that the atom is coupled to a finite band of modes with a $k$-independent coupling strength \cite{mahan}.

\section{Stationary states}\label{sec-ss}

The spectrum of the system described above consists of  a continuum of {\it unbound} stationary states, associated with a finite band of energies, and a pair of {\it bound} states corresponding to a pair of discrete levels. Either type, especially the former, has been studied in the literature \cite{scattering,sun,cinesi, longo,tureci}. 
Here we show how it is possible to derive at the same time both bound and unbound states through the Green function approach \cite{economou}. 

The Green function is defined in terms of Hamiltonian $\hat H$ as $\hat{G}(z)\ug({z\meno\hat{H}})^{-1}$, where $z$ is a complex variable. Unbound and bound stationary states correspond to branch cuts and poles of the Green function \cite{economou}. As shown in detail in Appendix \ref{app-green}, in our case the Green function takes the form 

\begin{eqnarray}  \label{green}
\hat{G}(z)\ug\hat{G}_0(z)\!\left\{\openone\piu\left[f\,|e\rangle \langle 0|\piu{\rm H.c.}\piu f_{1}\,|e\rangle \langle e|\piu f_2\,|0\rangle \langle 0| \,\right]  \!\hat{G}_0(z)\right\}\,,
\end{eqnarray}
where $\hat G_0(z)$ is the bare Green function associated with $\hat H_0$ [see \eq(\ref{H0})], $f(z)$ is the complex function given by
\begin{eqnarray}
\label{green0}
f(z)=\frac{g}{1-g^2G_{0e}(z)G_{00}(z)}\,\label{fz}
\end{eqnarray}
while $f_1(z)\ug g G_{00}(z) f(z)$, $f_2(z)\ug g G_{0e}(z) f(z)$
with $G_{0j}(z)=\langle j| \hat{G}_0(z)| j\rangle$ for $j\ug e,0$. It turns out (see Appendix  \ref{app-green}) that $G_{0e}(z)\ug z^{-1}$ and, in the thermodynamic limit $N\!\gg\!1$,
\begin{eqnarray}
G_{00}(z)=\frac{1}{2\pi}\!\int_{-\pi}^{\pi}\!\!dk\,\frac{1}{z-2 J \cos k}\,.\label{G00}
\end{eqnarray}
$G_{00}(z)$ has a branch cut on the real axis for $-2J\!\le\!z\!\le\!2J$ and hence so does $\hat G(z)$.
Instead, at any $z$ not coinciding with this singular line function $G_{00}(z)$ can be worked out as [\cf\eq(\ref{int-out}) in Appendix \ref{app-int}]
\begin{eqnarray}
G_{00}(z)\ug\frac{1}{\sqrt{z^2-4 J^2}} \,\,\,\,\,\,\mbox{for\,\,\,\,} z\notin[-2J,2J]\,\,.\label{G00out}
\end{eqnarray}
In order to find the poles of $\hat G(z)$, in virtue of \eq(\ref{fz}) we need to find the roots of equation
$1\meno g^2 G_{0e}(z) G_{00}(z)=0$ within the domain $z\notin[-2J,2J]$. Recalling that $G_{0e}(z)\ug z^{-1}$ and using \eq(\ref{G00out}), the above equation takes the form $1\meno \frac{g^2}{z\sqrt{z^2-4J^2}}\ug0$.

To summarize,
\begin{description}
\item $\!\hat{G}(z)$ has poles for $z\!\notin\![-2J,2J]$ fulfilling $1\meno \frac{g^2}{z\sqrt{z^2-4J^2}}\ug0\,;$
\item $\!\hat{G}_0(z)$ has a branch cut for $z\in[-2J,2J]$, and so does $\hat{G}(z)$\,.
\end{description}

\subsection{Discrete energies and bound states}

$\hat G(z)$ has two poles on the real axis at $z\ug \omega_{\pm}$, calculated as the two real roots of the equation $1\meno \frac{g^2}{z\sqrt{z^2-4J^2}}\ug0$ 
\begin{equation} \label{discrete}
\omega_{\pm}\ug\pm\sqrt{2J^2+\sqrt{g^4+4J^4}}\,.
\end{equation}
Such poles are also the discrete eigenvalues of $\hat H$ in the single-excitation sector. As expected, for any finite $g$, $|\omega_{\pm}|\!>\!2J$, i.e., these fall out of the continuous band (branch cut). In the weak-coupling (or equivalently strong-hopping) limit $g\!\ll\!J$, $\omega_\pm\!\simeq\!\pm2J$, i.e., the two levels collapse on the band edges. In the strong-coupling limit $g\!\gg\!J$, instead, they reduce to $\omega_\pm\!\simeq\!\pm g$ since we retrieve a standard Jaynes-Cummings model where the atom significantly interacts only with the 0th cavity. In passing, note that on a strictly mathematical ground the two discrete levels appear at any finite $g$. \\
Next, we derive the stationary states $|\Psi_\pm\rangle$ associated with the energies $\omega_{\pm}$, i.e., obeying the eigenvalue equation $\hat H |\Psi_\pm\rangle\ug\omega_{\pm}|\Psi_\pm\rangle$. According to the Green function theory \cite{economou}, the associated projectors $|\Psi_\pm\rangle\langle\Psi_ \pm|$ are the residue of $\hat G(z)$ at $z\ug \omega_{\pm}$. We calculate them in 
Appendix \ref{app-bound} and find
 \begin{eqnarray}\label{psipm}
 |\Psi_\pm\rangle\ug \pm\mathcal{N}\!\sum_{x}(\pm\,\varrho)^{|x|}|x\rangle + \sqrt{p_{b}}\,|e\rangle\,
 \end{eqnarray}
 with
  \begin{eqnarray}
 \mathcal{N}&\ug&\sqrt{ \frac{(1-p_{b})(1-\varrho^2)}{(1+\varrho^2)}}\label{N}\,,\\
 \varrho&\ug&\frac{\omega_{+}-\sqrt{\omega_+^2-4J^2}}{2J} \label{rho}\,,\\
p_b&\ug&\frac{g^4}{2\omega_+^2\,(\omega_+^2-2J^2)} \ug \frac{\eta^4}{2\left(\eta^4+2 \sqrt{\eta^4+4}+4\right)} \label{pb}\,,
 \end{eqnarray}
 where we have introduced the rescaled coupling strength
 \begin{eqnarray}
\eta\ug\frac{g}{J}\,\,.\label{eta}
\end{eqnarray}
The exponential decay of the photon amplitude away from $0$th cavity ( note that $\varrho \!< \!1$) confirms that $|\Psi_\pm\rangle$ is a {\it bound} stationary state. This justifies our notation for $p_b$, namely the probability to find the atom in $|e\rangle$ when the system is in either of the two bound states. 

\subsection{Continuous spectrum and unbound states}

As discussed above, the branch cut of $\hat G(z)$ coincides with the continuous spectrum of energies of $\hat H$, which is the band $[-2J,2J]$. As the same holds for $\hat H_0$, we call $\omega_k \ug 2J \cos k$ with $-\pi\!\le\!k\!\le\!\pi$ an arbitrary eigenvalue of $\hat H$ within the band. The corresponding eigenstate $|\Psi_k\rangle$, according to the Green function theory \cite{economou}, has to be chosen from the pair of states
\begin{equation}\label{psik}
  |\Psi_{k}^{\pm}\rangle=|\varphi_{k}\rangle \piu \hat{G}^{\pm}(\omega_{k})\hat{H_{1}}|\varphi_{k}\rangle\,,
  \end{equation}
 where $\hat{G}^{\pm}(\omega_{k})\ug\lim_{\delta \to 0^+} \hat{G}(\omega_{k}\!\pm\! i\delta)$. As shown in detail in Appendix \ref{app-band}, only the ``+" solution corresponds to the physical case where the photon is scattered from the atom, either reflected back  or transmitted forward. With this choice, the explicit form of $|\Psi_{k}\rangle$ reads 
 \begin{eqnarray}\label{psik} 
 |\Psi_{k}\rangle= \sum_{x} u_{kx}|x\rangle + u_{ke}|e\rangle\,
 \end{eqnarray}
with
 \begin{eqnarray}
u_{kx}&=&\frac{1}{\sqrt{N}}\, \left(e^{ikx}+{\gamma_{k}e^{i|kx|}}{}  \label{ukx}\right)\,,\\
 %u_{ke}&=&\frac{1}{\sqrt{N}}\frac{2 i  |{\sin k}|}{ 
 %  \left[4 i \left(\frac{J}{g}\right) |{\sin
%   k}|\cos k-\left(\frac{g}{J}\right)\right]}\,,
u_{ke}&=&\frac{1}{\sqrt{N}}\,\frac{2 \eta  |{\sin k}|}{ 
   4 \,  |{\sin
   k}|\cos k-i\eta^2}\,,\label{uke}\label{uke}
  \end{eqnarray} 
where $\gamma_k$ is given by
\begin{eqnarray}
% \gamma_{k}=\frac{1}{i4\left(\frac{J}{g}\right)^{2}|{\sin{k}}|\cos{k}-1}\label{Rk}\,.
 \gamma_{k}=-\frac{\eta^2}{4i\,|{\sin{k}}|\cos{k}+\eta^2}\label{Rk}\,.
\end{eqnarray}
Here, $\gamma_k$ and $1\piu\gamma_k$ represent the photon reflection and transmission probability amplitudes, respectively.

 \section{Atomic emission }
  \label{emission-sec}

We have now all the ingredients to  investigate the atomic emission into the field vacuum by an initially excited atom i.e., to study the time evolution of the state $|\Phi(0)\rangle \ug |e\rangle$. A straightforward decomposition of this in terms of all the bound and unbound stationary states as given by \eqs(\ref{psipm}) and (\ref{psik}), respectively,  leads to the following joint state at time $t\!\ge\!0$ 

\begin{equation}\label{psit} 
|\Phi(t)\rangle =\sum_{k}u_{ke}^* e^{-i\omega_{k}t} |\Psi_{k}\rangle \piu \sum_{\mu=\pm} \!  \sqrt{p_b} \,e^{-i\omega_{\mu}t}|\Psi_\mu\rangle\,.
\end{equation}

\subsection{Atom's excitation amplitude }

It is convenient to arrange the atom's excitation probability amplitude $\alpha(t)$ as the sum of two contributions as
\begin{equation}\label{at}
\alpha(t)\ug \langle e|\Phi(t)\rangle\ug\alpha_u(t)+\alpha_b(t)\,,
\end{equation}
where $\alpha_u(t)$  and $\alpha_b(t)$ are the contribution due to the unbound  and bound states respectively, i.e.,
\begin{eqnarray}
\alpha_u(t)&\ug&\langle e|\sum_{k}u_{ke}^*\,e^{-i\omega_{k}t} |\Psi_{k}\rangle\ug \sum_{k}|u_{ke}|^{2}e^{-i\omega_{k}t}\,,\label{aut}\\
\alpha_b(t)&\ug&\langle e|\sum_{\mu=\pm} \!  \sqrt{p_b} \,e^{-i\omega_{\mu}t}|\Psi_\mu\rangle\ug2p_b \cos(\omega_+ t)\,,\label{abt}
\end{eqnarray}
where we used $\omega_-\ug-\omega_+$ [\cf\eq\ref{discrete}].
With the help of \eq(\ref{uke}), for $N\!\gg\!1$, $\alpha_u(t)$ can be expressed in the integral form as
\begin{eqnarray}
\alpha_u(t)=\frac{\eta^2}{2 \pi} \!\int_{-\pi}^{\pi} \!\!dk \,\frac{\sin^2{k}}{\sin^2(2k)+\frac{\eta^4}{4}} \!\,\,e^{-i\omega_{k}t}\,.\label{aut2}
\end{eqnarray}

\subsection{Photon excitation amplitude }

Projection of \eq(\ref{psit}) onto $|x\rangle$ yields
\begin{eqnarray}\label{psixt}
\psi(x,t)\ug \langle x|\Phi(t)\rangle\ug \psi_u(x,t)\piu \psi_b(x,t)
\end{eqnarray}
with
\begin{eqnarray}\label{psiuxt}
\psi_u(x,t)&\ug&\!\sum_{k}\!u_{kx}u_{ke}^* e^{-i\omega_{k}t}\,, \\
\psi_b(x,t)&\ug&\sqrt{p_b} \mathcal{N}\!\sum_{\mu=\pm} \mu  (\mu\,\varrho)^{|x|}e^{-i\omega_{\mu}t}\nonumber\,,\\
&\ug & 2 \sqrt{p_b} \mathcal{N}\varrho^{|x|} \left\{\begin{array}{c} 
    -i\sin\omega_{+}t \,\,\,\mbox{for\,\,\,\,} |x| \,\,\, {\rm even}\\
      \cos\omega_{+}t \,\,\,\,\,\,\,\,\mbox{for\,\,\,\,} |x| \,\,\,\, {\rm odd}
       \end{array} \right.
\,. \label{psibxt}
\end{eqnarray}
In analogy with the atomic amplitude, here $\psi_u(x,t)$ [$\psi_b(x,t)$] stands for the contribution from the unbound (bound) stationary states to the photon probability amplitude $\psi(x,t)$. With the help of \eqs(\ref{ukx}) and (\ref{uke}), for $N\!\gg\!1$, the former can be arranged in a wave-packet form as
\begin{equation}
 \psi_u(x,t)\ug \frac{\eta}{\pi}\!\!\int_{-\pi}^{\pi} \!\!dk\, \frac{e^{-i\omega_{k}t}\, |{\sin{k}}|}{i \eta^2 + 4|\sin{k}|\cos{k} }
 \left(  \!e^{i k x} \!-\! \frac{\eta^2 e^{i|kx |}}{\eta^2 \!+\! 4 i |\sin{k}|\cos{k} }\right).
 \label{psiuxt2}
 \end{equation}

\subsection{From exponential decay to Rabi oscillations}

In \figs2 and 3, we study the behavior of the atomic excitation $p_e(t)\ug|\alpha(t)|^2$ and photon probability distribution $p_x(t)\ug|\psi(x,t)|^2$, respectively, for different values of the rescaled coupling strength $\eta$ [see \eq(\ref{eta})]. The plots were drawn through numerical evaluation of integrals (\ref{aut2}) and (\ref{psiuxt2}).
\begin{figure}
\includegraphics[width=0.42\textwidth]{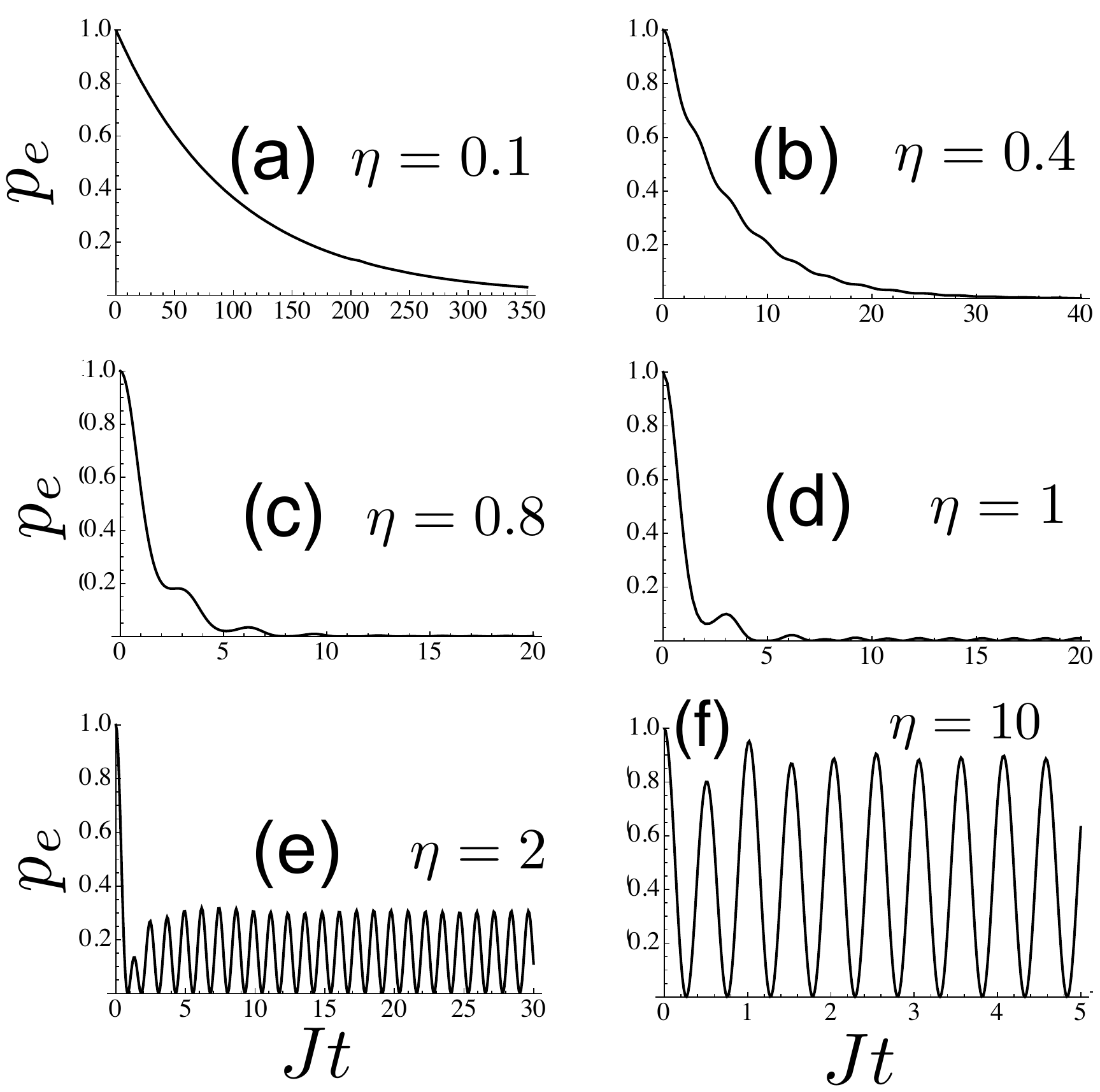}
\caption{Atom's excitation probability $p_e\ug|\alpha|^2$ against time (in units of $J^{-1}$) for increasing values of the rescaled coupling strength $\eta\ug g/J$. (a) $\eta\ug0.1$ . (b) $\eta\ug0.4$ . (c) $\eta\ug0.8$ . (d) $\eta\ug1$ . (e) $\eta\ug2$ . (f) $\eta\ug10$ .}
\end{figure} 

When the coupling strength is very low [\cf\fig2(a)] standard spontaneous emission  takes place and the atom excitation exhibits a purely exponential decay. In such conditions, indeed, ${\bf g\!\ll\!J}$ so the emitter does not sense the finiteness of the field band (correspondingly, the effective spectral density is flat). As the $g/J$ ratio is increased, secondary oscillations are introduced as shown in \figs2(b) and (c). As we discuss later, owing to the oscillatory term in \eq(\ref{abt}) $p_e(t\!\rightarrow\!\infty)$ never exactly vanishes for any finite $g$ (population trapping). Notwithstanding, as long as the ratio $g/J$ is not significantly larger than zero, for all practical purposes the atom in fact releases the entire amount of initial excitation to the field [in \fig2(c), e.g., $\eta\ug0.8$]. As $g$ further approaches $J$ [see \fig2(d) where $\eta\ug1$] the amount of excitation that remains trapped within the atom starts becoming more significant with $p_e(t)$ reducing to a stationary oscillation in the long-time limit. The amplitude of such stationary oscillation increases with $\eta$ until for $g\!\gg\!J$, namely $\eta\!\gg\!1$, standard vacuum Rabi oscillations occur. In this limit, as opposed to the case where $g\!\ll\!J$, the free-field band ``seen" by the emitter has negligible width, hence an effective single-mode behavior takes place. 
One can easily identify three main regimes:\\
\begin{description}
\item (i) \,\,\,purely exponential decay;
\item (ii) \,non-exponential decay (showing secondary oscillations) with no significant population trapping;
\item (iii) significant population trapping giving rise to fractional decay.
\end{description}

\begin{figure*}
\begin{center}
\includegraphics[width=0.8\linewidth]{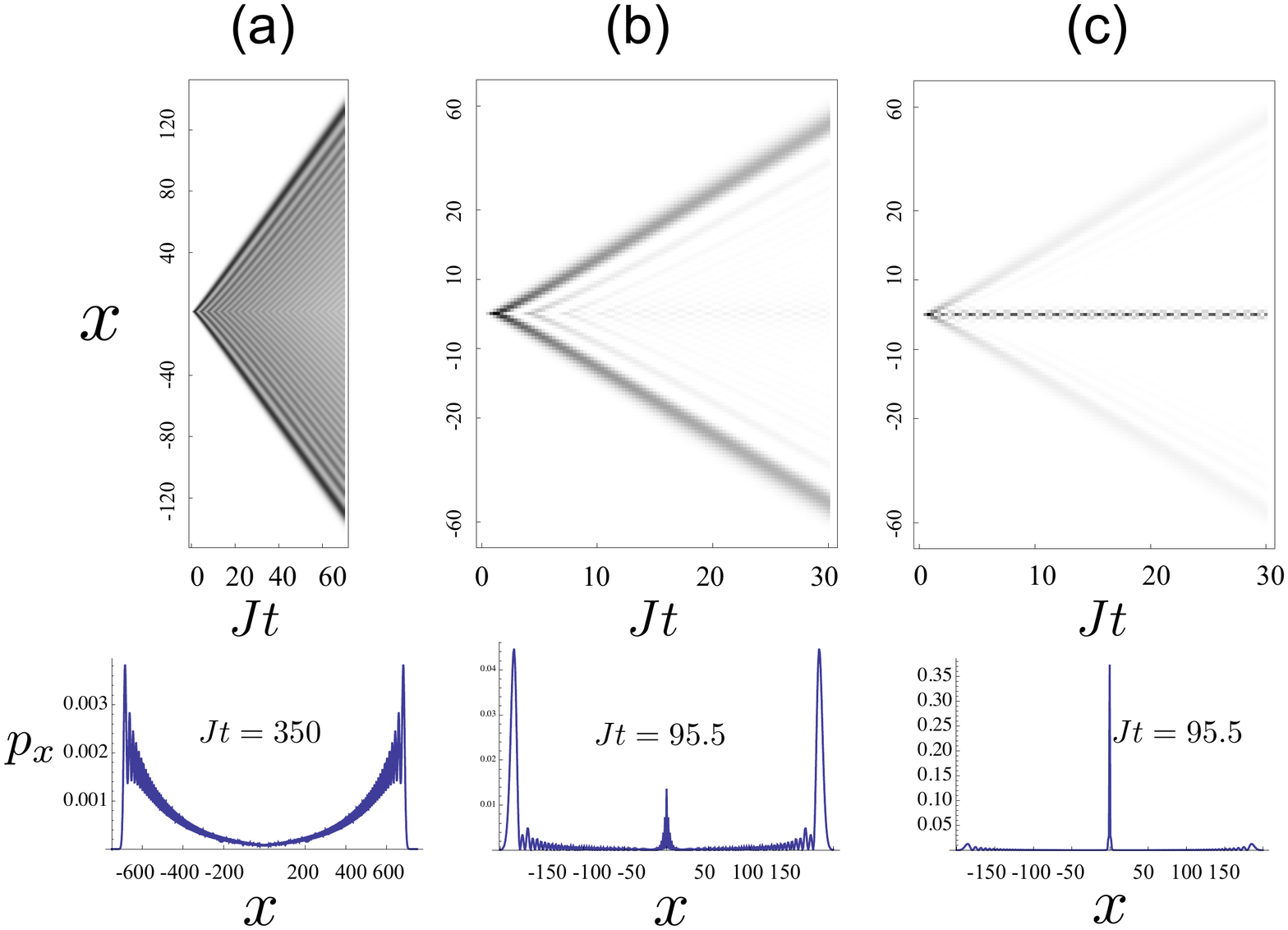}
\end{center}
\caption{(Color online) Space-time diagram of the photon probability distribution function $p_x(t)$ along the cavity array (top panels) for $\eta=0.1$ (a), $\eta=0.8$ (b), $\eta=2$ (c). The bottom panels display $p_x$ at a very large time, specifically $Jt\ug350$ (a) and $Jt\ug95.5$ [(b) and (c)]. Throughout, time is measured in units of $J^{-1}$.}
\end{figure*}

In \fig3 we investigate the time evolution of the photon probability distribution along the cavity array for the three representative values $\eta=0.1$ (a), $\eta=0.8$ (b), $\eta=2$ (c)  corresponding to regimes (i), (ii) and (iii), respectively. For exponential decay [regime (i), \cf\fig2(a)], the photon wave function slowly spreads along the array as shown in \fig3(a) exhibiting a wiggled profile. At long enough times, the probability to find the photon within a finite region around the atom vanishes [see bottom panel in \fig3(a) which shows $p_x$ at a very large time]. In the opposite regime [see \fig3(c)], when population trapping takes place [regime (iii), \cf\figs2(e)-(f)] most of the photon wave function remains localized around the atom [see also bottom panel]. Such localised light is periodically absorbed and next reemitted by the atom as is clear from the top-panel diagram.
While a detailed discussion of regime (ii) is postponed to the next section, here we discuss in more quantitative terms the two limiting cases of exponential decay and Rabi oscillations.

\subsection{Exponential decay}

The time (real) function in \eq(\ref{aut2}) vanishes in the limit $t\!\rightarrow\!\infty$. This can be seen by noting first that in \eq(\ref{aut2}) we can replace $e^{-i\omega_{k}t}\!\rightarrow\!\cos({\omega_{k}t})$ since the contribution from the imaginary part vanishes (this yields and odd integrand function). This and  the Jacobi-Anger expansion \cite{abramowitz}
\begin{eqnarray}
\cos(z \cos k)\ug J_0(z)\piu2\sum_{n=1}^\infty(-1)^n J_{2n}(z)\cos(2nk)
\end{eqnarray}
allow to express $\alpha_u(t)$ as the series of Bessel functions
\begin{eqnarray}
\alpha_u(t)\ug a_0\, J_0(2Jt)\piu\sum_{n=1}^\infty a_n \,J_{2n}(2Jt)\,
\end{eqnarray}
with $a_0\ug I_0$, $a_{n\ge1}\ug2(-1)^n I_n$ and where $I_n$ is obtained from integral (\ref{aut2}) by replacing the complex exponential with $\cos(2nk)$. Due to the appearance of the Bessel functions, function $\alpha_u(t)$ eventually decays to zero after exhibiting  secondary oscillations, these becoming less and less significant as $\eta\!\rightarrow\!0$. When $\eta$ is very small (i.e., hopping is very strong) a pure exponential decay arises
\begin{eqnarray}
\alpha_u(t)\simeq e^{-\frac{\eta^2}{2}Jt}\,\,\,\,{\rm for}\,\,\,\,\eta \ll 1\,, \label{lindblad}
\end{eqnarray}
a result that is proven in Appendix \ref{app-lindblad}. Correspondingly, $\alpha_b(t)\!\simeq\!0$ [\cf\eq(\ref{pb})].

\subsection{Rabi oscillations}

In contrast to $\alpha_u(t)$, the contribution due to the bound states \eq(\ref{at}) is always a pure oscillation at frequency $\omega_+$ [\cf\eq(\ref{discrete})], the amplitude of which according to \eq(\ref{pb}) ranges from 0 (for $\eta\!\ll\!1$) to 1 (for $\eta\!\gg\!1$). In the latter case, $\omega_+\!\simeq\! g$ [see discussion following \eq(\ref{discrete})] while $\alpha_u(t)\!\simeq\!0$ [\cf\eq(\ref{aut})] and hence $\alpha(t)\!\simeq\!\alpha_b(t)\!\simeq\! \cos (g t)$: as expected, we retrieve vacuum Rabi oscillations corresponding to the coherent interaction between the atom and the 0th cavity. In this limit, the unbound-states contribution to the field [\cf\eqs(\ref{psixt}) and (\ref{psiuxt2})] becomes negligible, hence $\psi(x,t)\!\simeq\! \psi_b(x,t)$ at any $x$ and $t$. In the strong-coupling limit, $\psi_b(x,t)$ becomes strongly peaked around the central cavity $x\ug0$, whose corresponding probability amplitude oscillates as $\!\sin(\omega_+t)\!\simeq\!\sin(gt)$ according to \eq(\ref{psibxt}). In such a limit, most part of the field is concentrated next to the atom and a continuous atom-field energy exchange goes on at angular frequency $2\omega_+$. Such limiting behaviour is already evident from \fig3(c) where $\eta\ug2$, showing that the field is distributed in space mostly within the central cavity where it exhibits cyclic modulation of its intensity.

\section{Photon localisation without fractional decay} 
\label{study}
In regime (ii), as shown in \fig3(b), despite negligible population trapping and hence in the presence of full atomic  decay [\cf\fig2(c)] a small yet appreciable fraction of the photon wave function remains localized in the neighborhood of the atom [central region of the top-panel diagram]. This is confirmed by the field profile at a very long time (bottom panel): although most of the emitted light departs away from the emitter, a significant photon density survives indefinitely next to the atom.
To gain a better insight of the dynamics in such regime let us rearrange $\psi_b(x,t)$ in an exponential form as
\begin{eqnarray}\label{psibxt2}
\psi_b(x,t)\ug A\, {\rm e}^{\frac{-|x|}{\lambda}}\chi(x,t)\,,
\end{eqnarray}
where $\chi(t)\ug -i\sin(\omega_+t)$ [$\chi(t)\ug \cos(\omega_+t)$] for even (odd) $|x|$ (we omit the spatial dependance of $\chi$ for simplicity) while
\begin{eqnarray}
A&\ug&\frac{\eta^5}{\left(\eta^4+2
   \sqrt{\eta^4+4}+4\right)\sqrt{\sqrt{\eta^4+4}-2} }\,,\label{Aloc}\\
\lambda&\ug&\frac{1}{\log \left[\frac{1}{2}
   \left(\sqrt{\sqrt{\eta^4+4}-2}+\sqrt{\sqrt{\eta^4+4}+2
   }\right)\right]}\label{lambda}\,,
\end{eqnarray}
where we have used \eqs(\ref{N})-(\ref{pb}).
\begin{figure}
\includegraphics[width=0.35\textwidth]{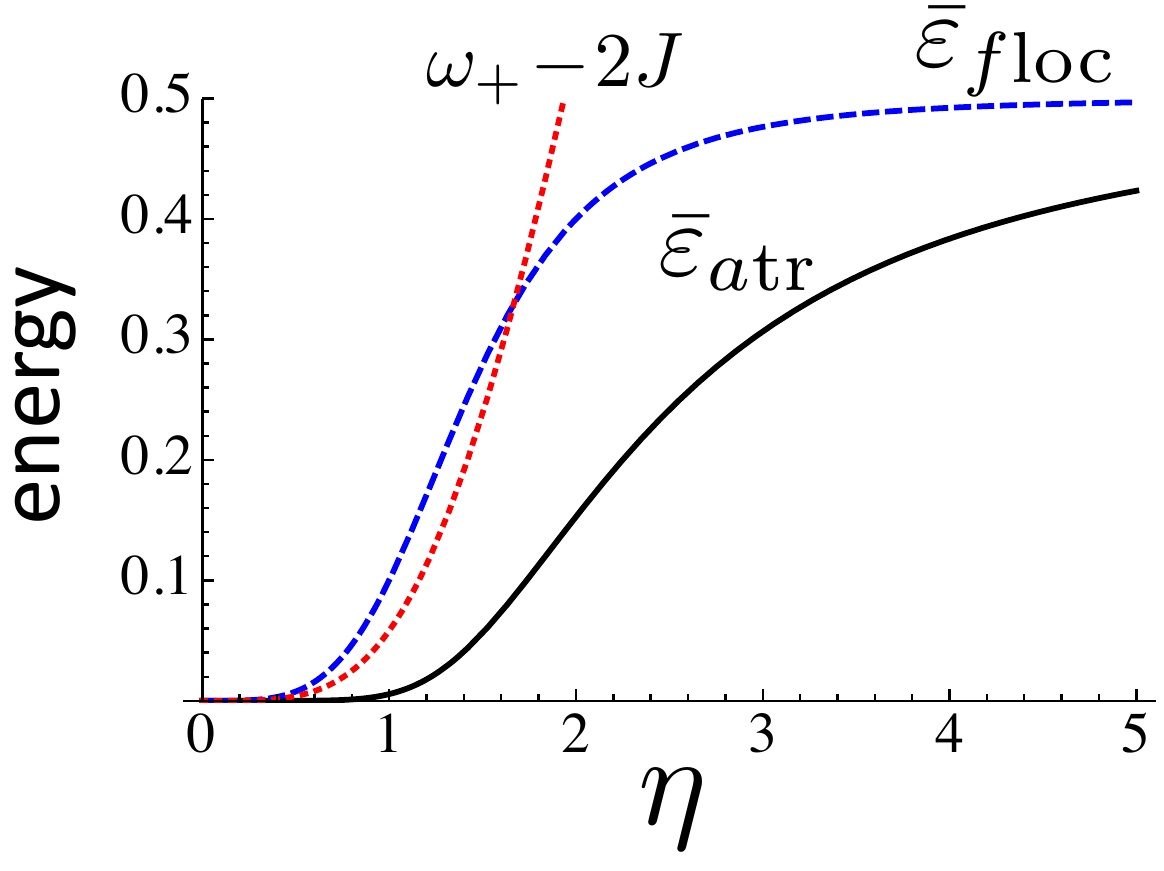}
\caption{(Color online){ Time-averaged trapped mean atomic energy $\bar{\varepsilon}_{a{\rm tr}}$ (black solid line) and localized mean field energy $\bar{\varepsilon}_{f{\rm loc}}$ (blue dashed) against $\eta\ug g/J$. For comparison, we also plot the behavior of $\omega_+\!-\!2J$ (in units of $J)$, as given by \eq(\ref{discrete}), which measures the occurrence of bound states. $\bar{\varepsilon}_{f{\rm loc}}$ and $\bar{\varepsilon}_{a{\rm tr}}$ are in units of the atomic frequency.}  
 \label{fig_loc}}
\end{figure} 
\eq(\ref{psibxt2}) accounts for field localization and should be analyzed in combination with the stationary atomic oscillations described by \eq(\ref{abt}). 
As excitation can be trapped around the atom in either form (photonic or excitonic) it is natural to assess the relative importance of each of the two contributions. At long enough times,  $\psi(x,t)\!\rightarrow\!\psi_{b}(x,t)$ for $|x|\!\lesssim\!\lambda$, while $\alpha(t)\!\rightarrow\!\alpha_b\cos(\omega_+ t)$. It is convenient to define (at long times) the time-averaged localized mean photon energy as $\bar{\varepsilon}_{f{\rm loc}}\ug\sum_x\overline{| \psi_b(x,t)|^2}$ (in units of the atomic frequency). This is equivalent to the time-averaged probability that a photon is found at $|x|\!\lesssim\!\lambda$ at long times. Using \eqs(\ref{psibxt2})-(\ref{lambda}), we compute
\begin{eqnarray}\label{epsfloc}
\bar{\varepsilon}_{f{\rm loc}}&\ug&\!\sum_{x=-N/2}^{N/2-1}\left(A{\rm e}^{-\frac{|x|}{\lambda}}\right)^2\overline{|\chi(t)|^2}\ug \frac{A^2}{2}\!\! \sum_{x=-N/2}^{N/2-1}\!\!\!{\rm e}^{-\frac{2|x|}{\lambda}}\nonumber\\
&&\stackrel{N\gg{1}}{\longrightarrow} \frac{A^2\coth{\lambda^{-1}}}{2}\ug\frac{\eta^4}{2 \eta^4+8}\,,
\end{eqnarray}
where the $1/2$ ratio derives from the temporal average of $\sin^2(\omega_+ t)$ and $\cos^2(\omega_+ t)$ entering function $\chi(t)$. Correspondingly, we define (at long times) the time-averaged trapped mean atomic energy as $\overline{P}_{e{\rm tr}}\ug\overline{|\alpha_b(t)|^2}$, which is explicitly expressed with the help of \eqs(\ref{pb}) and (\ref{abt}) as
\begin{eqnarray}\label{epsatr}
\bar{\varepsilon}_{a{\rm tr}}\ug {\overline{|\alpha_b(t)|^2}}\ug2p_b^2\ug\frac{\eta^8}{2\left(\eta^4+2 \sqrt{\eta^4+4}+4\right)^2}\,.
\end{eqnarray}
A major feature arising from the \eqs(\ref{epsfloc}) and (\ref{epsatr}) is the different functional behaviour of $\bar{\varepsilon}_{f{\rm loc}}$ and $\bar{\varepsilon}_{a{\rm tr}}$. In particular, for small $\eta$, $\bar{\varepsilon}_{f{\rm loc}}\!\sim\!\eta^4$ while $\bar{\varepsilon}_{a{\rm tr}}\!\sim\!\eta^8$. Such different scaling behaviour is evident in \fig4, where we plot $\bar{\varepsilon}_{f{\rm loc}}$ and $\bar{\varepsilon}_{a{\rm tr}}$ against $\eta$. Either function vanishes at the origin and saturates to 1/2 for $\eta\!\gg\!1$ corresponding to the expected behaviours in regime (i) and (iii), respectively. Because of the different scaling law, though, their behaviours at intermediate values are quite different. Despite the absence of a mathematical threshold, either function features a physical threshold beyond which it is significantly different from zero. These thresholds are $\eta\!\simeq\!0.4$ for $\bar{\varepsilon}_{f{\rm loc}}$ and $\eta\!\simeq\!0.9$ for $\bar{\varepsilon}_{a{\rm tr}}$. Moreover, compared to $\bar{\varepsilon}_{f{\rm loc}}$, $\bar{\varepsilon}_{a{\rm tr}}$ converges to 1/2 quite slowly. This explains why as $\eta$ grows from zero, field localization becomes significant before population trapping.
\begin{figure}
\includegraphics[width=0.35\textwidth]{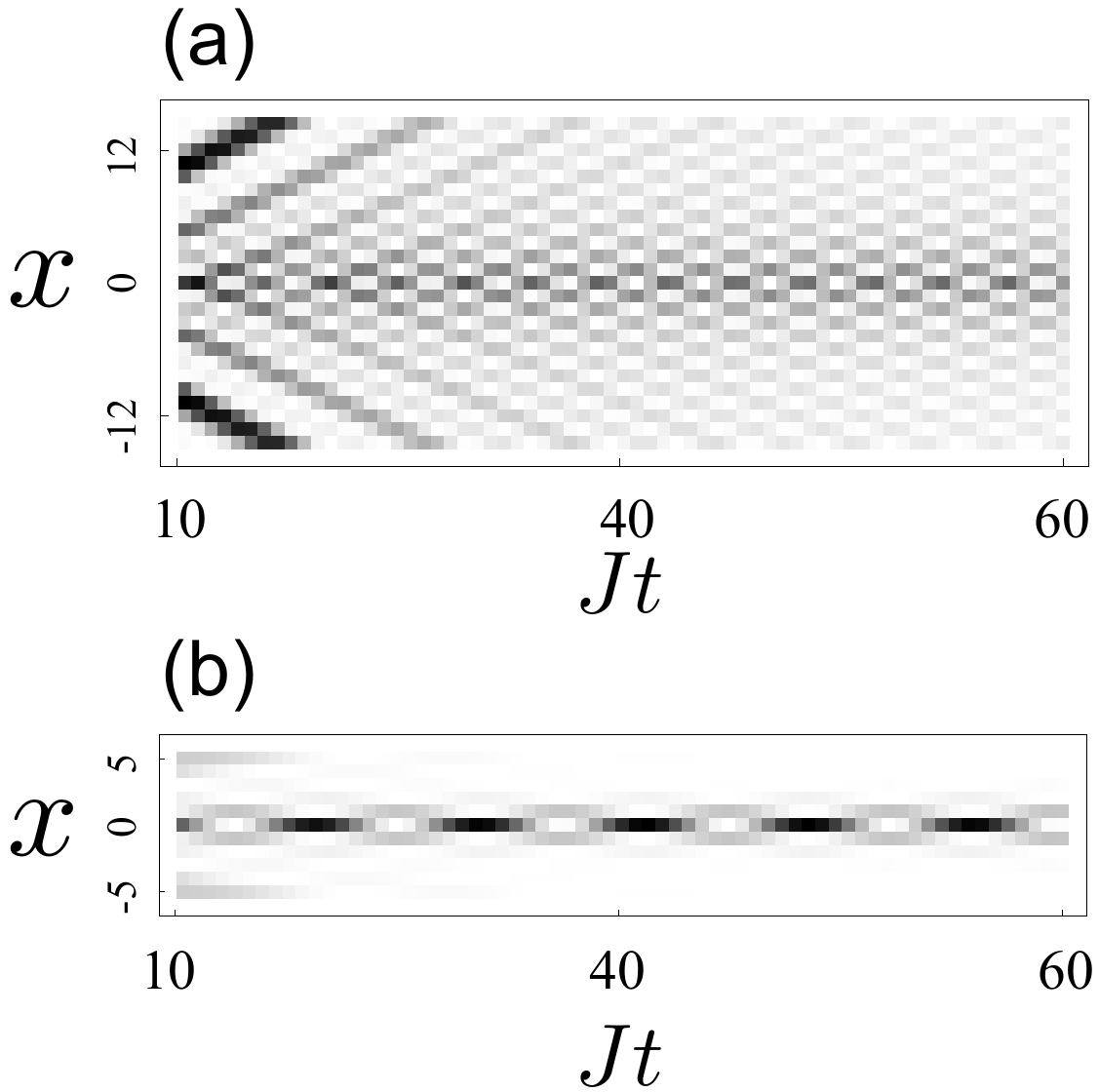}
\caption{(Color online){ Space-time diagram of the photon probability distribution function $p_x(t)$ in the vicinity of $x\ug0$ for $\eta\ug0.8$ (a) and $\eta\ug2$ (b). Time is measured in units of $J^{-1}$.} }
\end{figure} 
In contrast to the energy exchange dynamics in regime (iii) for $g\!\gg\!J$, where -- as discussed -- a full atom-field energy swap periodically occurs, \fig4 shows that the averaged energy of the localized field in general exceeds the atomic one. Interestingly, this brings about that for values of $\eta$ between the two thresholds, i.e., $0.4\!\lesssim\!\eta\!\lesssim\!0.9$, {\it at long times} only a negligible fraction of the localized field energy  is periodically returned to the atom. As a consequence, within this range, the time-oscillating profile of the localized field entailed by the sinusoidal functions of $\omega_+ t$ [featured in $\chi(t)$] is essentially due to an energy redistribution of the field among the cavities next to $x\ug0$. This behaviour can be appreciated through a closer inspection of the central region around $x\ug0$ in the space-time diagrams in \figs3(b) and (c), which we display on a proper scale in \figs5(a) and (b), respectively. In the latter case, as remarked in the previous section, the overall field intensity exhibits a periodic modulation (due to full periodic absorption of the atom). In the former case, instead, the overall intensity is about constant in time but periodically undergoes a substantial redistribution in space as witnessed by the long-time checkerboard-like pattern in \fig5(a).

\section{Conclusions}\label{concl}

In this paper, we have investigated the excitation transport dynamics in the emission process of an atom into a long CCA By studying the time evolution of the atomic population as well as the photon probability distribution function across the array, we have highlighted the occurrence of three regimes. For very weak values of the coupling strength, standard exponential decay of the atom's population takes place. Simultaneously, the photon wave function spreads in either direction of the array in such a way that, for long enough times, it vanishes within any finite region around the central cavity. In contrast, when the coupling is strong, the field fully localizes next to the atom's site since in this limit the dynamics reduces to the well-known Jaynes Cummings model. Correspondingly, full population trapping takes place in that the atom periodically exchanges the entire content of its initial energy with the localized field so as to exhibit standard vacuum Rabi oscillations. At intermediate values of the coupling rate -- in general -- the dynamics features partial field localization and atomic fractional decay, the latter being due to population trapping that manifests as a residual oscillation of the atom's excitation at large times. While significant population trapping is always accompanied by significant field localization, {\bf we found that} the converse is not true. This is due to the different functional dependence of such two phenomena on the coupling rate. Specifically, significant field localization occurs beyond a {\bf physical} threshold value which is lower than the population trapping one. In the region between the two thresholds -- at long times -- the localized field periodically undergoes a mere spatial redistribution in the vicinity of the atom without returning energy to it. Such behaviour, as we have shown, arises from the peculiar properties of the two bound stationary states of the atom-field joint system. Between the two aforementioned thresholds, each bound state features a negligible excitonic component.

As discussed in Section \ref{sec-h}, the one-atom CCA addressed here is equivalent to the tight-binding model represented by the network in \fig1(b). Similar Hamiltonians have been analyzed mostly in the framework of optical waveguides \cite{physrep}. One such case is that of a semi-infinite linear chain with one hopping rate differing from all others \cite{longhi-primosito}, which can be mapped into a one-atom CCA where the atom is coupled to a boundary cavity. Another instance is a linear chain featuring a central site which is coupled at a rate $g$ with the two nearest-neighbour  sites, all the remaining hopping rates of the chain being equal to $J$ \cite{longhi-trompeter,nota-threshold}. We have numerically assessed that, in the case of such defect models, field localization is always accompanied by significant population trapping. In this respect, the distinctive properties of the model addressed here are under current investigations \cite{nota-epjb}.

\subsection*{Acknowledgements}

We are grateful to D. Burgarth for invaluable discussions. We acknowledge support from MIUR (PRIN 2010 - 2011)

\appendix
\section{Green function}\label{app-green}

Let $\hat G_0(z)$ be the green function associated with $\hat H_0$ [see \eq(\ref{H0})]. The eigenstates of the latter, in the overall Hilbert space, are $\{|\varphi_k\rangle\}$ and $|e\rangle\langle e|$ with eigenvalues $\omega_k\ug 2J \cos k$ and 0, respectively [see \eqs(\ref{omegak}) and (\ref{phik})]. Hence, using the corresponding spectral decomposition of $\hat H_0$,
\begin{equation}\label{G0}
\hat{G}_0(z)=\frac{|e\rangle\langle e|}{z}+\sum_{k}\frac{|\varphi_{k}\rangle\langle\varphi_{k}|}{z-\omega_k}\,.
\end{equation}
$\hat G (z)$ can be linked with $\hat G_0(z)$ through the series expansion \cite{economou}
\begin{equation}\label{G-G0}
\!\hat{G}(z)\ug\hat{G}_0(z)+\hat{G}_0(z)\hat{H}_1\hat{G}_0(z)+\hat{G}_0(z)\hat{H}_1\hat{G}_0(z)\hat H_1 \hat G_0(z)+...
\end{equation}
This can be regarded as a series expansion of $\hat G(z)$ in terms of powers of the coupling strength $g$, the form of which reads (we omit the dependence on $z$ since this is unnecessary for the scopes of this section)
\begin{eqnarray}
\hat G\ug \hat G^{(0)}\piu \hat G^{(1)} g\piu  \hat G^{(2)}g^2\piu...\,,
\end{eqnarray}
where $\hat G^{(0)}\!\equiv\!\hat G_0$, $\hat G^{(1)}\ug g^{-1}{\hat G_0 \hat H_1\hat G_0},...$, each $\hat G^{(k)}$ thus being independent of $g$.
Using this along with \eq(\ref{G0}) and the fact that in the one-excitation-sector $\hat H_1\ug g |e\rangle \langle 0|\piu{\rm H.c.}$ [\cf\eq(\ref{H1})], up to the 4th power in $g$ we find
\begin{eqnarray}
\hat{G}^{(1)}&\ug&\hat{G_{0}}\left(|e\rangle\langle 0| \piu|0\rangle\langle e| \right)\hat{G_{0}}\,,\nonumber\\
\hat{G}^{(2)}&\ug&\hat{G_{0}}\left(G_{00}|e\rangle\langle e| \piu G_{0e} |0\rangle\langle 0| \right)\hat{G_{0}} \,,\nonumber\\
\hat{G}^{(3)}&\ug&G_{00} G_{0e} \,\,\hat{G_{0}}  \!\left( |e\rangle\langle 0| \piu |0\rangle\langle e| \right)\hat{G }_0 \,,\nonumber\\
\hat{G}^{(4)}&\ug&G_{00} G_{0e} \,\hat{G_{0}}\! \left(G_{00} |e\rangle\langle e| \piu G_{0e}  |0\rangle\langle 0| \right)\hat{G }_0 \ \nonumber
\end{eqnarray}
By induction, this is generalized for any integer $k\!\ge\!1$ as
\begin{eqnarray}
\hat{G}^{(2k+1)}&\ug& (G_{00}  G_{0e} )^{k}\, \hat{G_{0}}\! \left(  |e\rangle\langle 0| \piu |0\rangle\langle e| \right)\hat{G_{0}} \,,\nonumber\\
\hat{G}^{(2k)}&\ug&  (G_{00}  G_{0e} )^{k-1}\, \hat{G_{0}} \!\left(G_{00} |e\rangle\langle e| \piu G_{0e}  |0\rangle\langle 0|\right)\hat{G_{0}}  \,.\nonumber
\end{eqnarray}
Note that even and odd power terms are always proportional to $\hat{G_{0}}\left(G_{00} |e\rangle\langle e| \piu G_{0e}  |0\rangle\langle 0| \right)\hat{G_{0}}$ and $\hat{G_{0}}\left( |e\rangle\langle 0| \piu|0\rangle\langle e| \right)\hat{G_{0}}$, respectively. By introducing now the geometric series $f\ug g\sum_{n=0}^{\infty}(g^2 G_{00}  G_{0e})^{n}$, whose sum coincides with \eq(\ref{fz}), together with functions $f_1(z)\ug g G_{00}(z) f(z)$ and $f_2(z)\ug g G_{0e}(z) f(z)$ $f_1$ and $f_2$, we straightforwardly end up with \eq(\ref{green}). From \eq(\ref{G0}), $G_{0e}(z)\ug z^{-1}$ and
\begin{eqnarray}
G_{00}(z)=\sum_{k}\frac{|\langle\varphi_k|0\rangle|^2}{z-\omega_k}\,\,\stackrel{N\gg{1}}{\longrightarrow} \,\, \frac{1}{2\pi}\!\int_{-\pi}^{\pi}\!\!dk\,\frac{1}{z-2 J \cos k}\,,
  \end{eqnarray}
 where we used \eqs(\ref{omegak}) and (\ref{phik}) and computed the thermodynamic limit $N\!\gg\!1$ [in this limit, owing to \eq(\ref{k}), $2\pi/N\!\rightarrow\!dk$].

\section{Useful integrals}\label{app-int}

Let $j$ be an integer number, $A$ a positive constant and $z$ a complex variable. Then \cite{economou}
\begin{eqnarray} 
\frac{1}{2\pi}\!\!\int_{-\pi}^{\pi}\!\!\!dk\frac{e^{ikj}}{(z-2A\cos k)}&\ug\!&\frac{\left({\tilde z}\meno\sqrt{\tilde{z}^2\meno1}\right)^{|j|}}{\sqrt{z^2\meno4A^2}}\,\,{\rm for}\,\,\,\tilde z\!\notin\![-1,1]\label{int-out}\,,\,\,\,\,\,\,\,\,\,\\
\frac{1}{2\pi}\!\!\int_{-\pi}^{\pi}\!\!\!dk\frac{e^{ikj}}{(z-2A\cos k)}&\ug\!&\frac{\mp i\left({\tilde z}\!\mp\!\sqrt{1\meno\tilde{z}^2}\right)^{|j|}}{\sqrt{4A^2-z^2}}\,\,{\rm for}\,\,\,\tilde z\!\in\![-1,1]\label{int-in}\,,\,\,\,\,\,\,\,\,\,\,\,\,
 \end{eqnarray}
 where
 \begin{eqnarray} 
 \tilde z=\frac{z}{2A}\label{ztilde}\,.
 \end{eqnarray}
 The double sign in \eq(\ref{int-in}) arises from the replacement $z\!\rightarrow\!z\!\pm\!i\delta$ followed by the limit for $\delta\!\rightarrow\!0^+$.

\section{Bound stationary states} \label{app-bound}

Based on \eq(\ref{green}) and recalling that $f_1(z)\ug g G_{00}(z) f(z)$, $f_2(z)\ug g G_{0e}(z) f(z)$, the residues needed for the calculation of $|\Psi_\pm\rangle\langle \Psi_\pm|$ are given by
\begin{eqnarray} 
&&  r\ug {\rm Res}(f,\omega_{\pm})\ug\pm\frac{g^5}{2\sqrt{g^4+4J^4}\sqrt{2J^2+\sqrt{g^4+4J^4}}}\,,\,\,\,\,\label{resf} \\
&&   r_{1} \ug {\rm Res}(f_1,\omega_{\pm})\ug\frac{g^4}{2\sqrt{g^4+4J^4}}\,, \label{resf1}\\
&&  r_{2} \ug {\rm Res}(f_2,\omega_{\pm})\ug\frac{g^6}{2\sqrt{g^4+4J^4}{(2J^2+\sqrt{g^4+4J^4)}}}\,.\,\,\,\,\label{resf2}   
\end{eqnarray}
Substituting these in ${\rm Res}(\hat G,\omega_\pm)$  yields
\begin{eqnarray} \label{prpiu2}
|\Psi_\pm\rangle\langle \Psi_\pm| &\ug&\! \!-\frac{r}{2\pi\omega_{+}}\int_{-\pi}^{\pi}dk\,\frac{|e\rangle \langle\varphi_{k}| \piu{\rm H.c.}}{ \omega_k\mp\omega_{+}} \piu \frac{r_{1}}{\omega_+^2}{|e\rangle\langle e|} \nonumber\\ 
&&\piu \frac{r_{2}}{4\pi^2}\!\int_{-\pi}^{\pi}\!dk\!\int_{-\pi}^{\pi}\!dk'\frac{|\varphi_{k}\rangle\langle \varphi_{k'}|}{(\omega_k\mp\omega_{+})(\omega_{k'}\mp\omega_{+})}\,.\,\,\,\,\,\,\,
 \end{eqnarray}
The matrix elements of projector (\ref{prpiu2}) are calculated as
 \begin{eqnarray} 
\langle e|\Psi_\pm\rangle\langle \Psi_ \pm|e\rangle &\ug& \frac{r_{1}}{\omega_{+}^2}\,,\label{ee}\\
\langle x|\Psi_\pm \rangle\langle \Psi_ \pm|x'\rangle 
 &\ug& \frac{r_{2}}{4\pi^2}\!\!\int_{-\pi}^{\pi}\!\!dk\frac{e^{ikx}}{\omega_{+}\mp \omega_k}\!\!\int_{-\pi}^{\pi}\!\!dk' \frac{e^{-ik'x'}}{\omega_{+}\mp \omega_{k'}}\nonumber \\ 
&\ug& \frac{r_{2}}{\omega_{+}^2-4J^2}\,(\pm\,\varrho)^{|x|+|x'|}\,,\label{xx'}\\
\langle e|\Psi_\pm\rangle\langle \Psi_ \pm|x\rangle& \ug& \langle x|\Psi_\pm\rangle\langle \Psi_ \pm|e\rangle\ug\pm\frac{r}{\omega_{+}\sqrt{\omega_{+}^2-4J^2}}\,(\pm\,\varrho)^{|x|}\,, \label{ex}\nonumber\\
 &&
 \end{eqnarray} 
where $\varrho$ is given in \eq(\ref{rho}).
In deriving \eqs(\ref{xx'}) and (\ref{ex}) we have used \eq(\ref{int-out}) in Appendix \ref{app-int}.

Based on \eqs(\ref{rho}), (\ref{resf2}) and (\ref{ee}), the following identity holds
\begin{eqnarray} \label{identity}
\frac{r_2}{\omega_+^2-4J^2}\ug \frac{(1-p_b)(1-\varrho^2)}{1+\varrho^2}   \,.
\end{eqnarray}
Using this and the identity $\sqrt{r_1 r_2}\ug r$ [\cf\eqs(\ref{resf})-(\ref{resf2})] one can check that the projector associated with state (\ref{psipm}) has the same matrix elements as those in \eqs(\ref{ee}), (\ref{xx'}) and (\ref{ex}). This proves (up to an irrelevant phase factor) that  the state corresponding to projector (\ref{prpiu2}) is indeed given by \eq(\ref{psipm}).

\section{Unbound stationary states} \label{app-band}

 The second term on the right-hand side is the perturbation of $|\varphi_{k}\rangle$ due to the atom-field coupling. This is calculated as
 \begin{equation}\label{gpmh1}
{\hat{G}^{\pm}(\omega_{k})\hat{H_{1}}|\varphi_{k}\rangle}\ug\!\frac{g\,}{\sqrt{N}}\left[\left(1 \piu\! \frac{f_{1}(\omega_{k}^{\pm})}{\omega_{k}^{\pm}}\right )\!\frac{|e\rangle}{\omega_{k}^{\pm}} \piu \frac{f(\omega_{k}^{\pm})}{\sqrt{N}\omega_{k}^{\pm}}\!\sum_{k'}\!\frac{|\varphi_{k'}\rangle}{\omega_{k}^{\pm}-\omega_{k'}}\!\right],\nonumber
 \end{equation}
 where we set $\omega_{k}^{\pm}\ug\omega_{k}\!\pm\! i\delta$. 
Upon projection on $|x\rangle$, for $N\!\gg\!1$ we obtain
 \begin{equation}
\langle x|\hat{G}^{\pm}(\omega_{k})\hat{H_{1}}|\varphi_{k}\rangle \ug \frac{g}{\sqrt{N}}\!\frac{f(\omega_{k}^{\pm})}{2\pi}\!\!\int_{-\pi}^{\pi}\!\!dk' \frac{e^{ik'x}}{\omega_{k}^{\pm}\meno\omega_{k'}}\ug \frac{\gamma_{k\pm}}{\sqrt{N}}\varrho_{k\pm}^{|x|} \label{pert} 
 \end{equation}
 with
\begin{eqnarray}
\varrho_{k\pm}&=& \frac{\omega_{k}\pm\sqrt{\omega_{k}^{2}\meno4J^2}}{2J} \ug \cos{k}\pm i\,|{\sin{k}}|\,,\label{rhok}\\
 \gamma_{k\pm}&=&-\frac{1}{1\pm 4 i \left(\frac{J}{g}\right)^{2}|{\sin{k}}|\cos{k}}\label{Rk}\,,
\end{eqnarray}
where we used \eq(\ref{int-in}) in Appendix \ref{app-int}.
Based on \eqs(\ref{pert}) and (\ref{rhok}), we note that, for $0\!\le\!k\!\le\!\pi$, $\varrho_{k\pm}\ug\!e^{\pm ik}$ while $-\pi\!\le\!k\!\le\!0$ yields $\varrho_{k\pm}\ug\!e^{\mp ik}$. Hence, $\varrho_{k\pm}^{|x|}\ug\!e^{\pm i|k x|}$, which shows that the ``+" solution corresponds to the physical case where the photon is scattered from the atom, either reflected back  or transmitted forward [\eg, if $k\!>\!0$ $\rho_{k+}\ug e^{-ikx}$ ($\rho_{k+}\ug e^{ikx}$) for negative (positive) $x$]. In contrast, the ``-" solution does not correspond to a physically meaningful situation and we thus discard it. 

Projecting now \eq(\ref{gpmh1}) onto $|e\rangle$ yields
 \begin{equation}\label{egpmh1}
\langle e|\hat{G}^{\pm}(\omega_{k})\hat{H_{1}}|\varphi_{k}\rangle \ug  \frac{g}{\sqrt{N}\,\omega_k^\pm}\left[1\piu\frac{f_1(\omega_{k}^\pm)}{\omega_{k}^\pm}\right]\,,
 \end{equation}
 whose explicit form, compatible with the choise $\varrho_{k+}$, coincides with \eq(\ref{uke}).

\section{Proof of \eq(\ref{lindblad})}\label{app-lindblad}

Integral (\ref{aut2}) can be expressed as
\begin{eqnarray}
\alpha_u(t)\ug\frac{\eta^2}{ \pi} \!\!\int_{0}^{\pi} \!\!dk \,F(k) \!\,\,e^{-2 i J t\cos k}\label{aut3}
\end{eqnarray}
with
 \begin{eqnarray}
F(k)\ug\frac{\sin^2{k}}{\sin^2(2k)+\frac{\eta^4}{4}}\,,
\end{eqnarray}
where we have used that both $F(k)$ and the complex exponential in \eq(\ref{aut3}) [\cf\eq(\ref{omegak})] are even functions of $k$.
$F(k)$ is peaked around $k\ug\pi/2$, the height of the peak becoming infinite in the limit $\eta\!\rightarrow\!0$. Hence, for $\eta\!\ll\!1$, the dominant contribution to integral (\ref{aut3}) comes from values of $k$ close to $k\ug\pi/2$. One can therefore make the approximations $\sin k\!\simeq\!1$, $\cos k\!\simeq\!-k\piu\pi/2$, which yields $F(k)\!\simeq\!1/[4(k\meno\pi/2)^2\piu \eta^4/4]$, and, moreover, extend the integration range to $[-\infty,\infty]$. This entails
\begin{eqnarray}
\alpha_u(t)\!\simeq\!\frac{\eta^2}{ \pi} \!\!\int_{-\infty}^{\infty} \!\!dk \,\frac{e^{2 i J (k-\pi/2)t}}{4(k\meno\pi/2)^2\piu \eta^4/4} \!\equiv\! \frac{\eta^2}{ \pi} \!\!\int_{-\infty}^{\infty} \!\!dk \,\frac{e^{2 i J t k}}{4k^2\piu \eta^4/4}\,.\nonumber
\end{eqnarray}
This is proportional to the Fourier transform of a Lorentzian, which results in the exponential function in \eq(\ref{lindblad}).

\end{document}